\documentclass[preprint,12pt]{elsarticle}

\usepackage[T1]{fontenc}
\usepackage[utf8]{inputenc}
\usepackage{lmodern}
\usepackage{amsmath,amssymb,mathtools,bm}
\usepackage{booktabs,array,graphicx,multirow}
\usepackage{xcolor}
\usepackage{hyperref}
\usepackage{dsfont}
\usepackage{microtype}
\usepackage{xr}
\usepackage[nameinlink,noabbrev]{cleveref}

\hypersetup{
  colorlinks=true,
  linkcolor=blue!50!black,
  citecolor=blue!50!black,
  urlcolor=blue!50!black
}

\allowdisplaybreaks
\numberwithin{equation}{section}

\newcommand{\E}{\mathbb{E}}
\newcommand{\Prb}{\mathbb{P}}
\newcommand{\ind}{\mathds{1}}

\newcommand{\IF}{\mathrm{IF}}

\journal{Economics Letters}

\begin{document}

\begin{frontmatter}

\title{Doubly Robust Local Projections Difference-in-Differences}

\author[ufpel]{Daniel de Abreu Pereira Uhr\corref{cor1}}
\ead{daniel.uhr@gmail.com}
\author[ufsc]{Guilherme Valle Moura}
\ead{gvallemoura@gmail.com}
\cortext[cor1]{Corresponding author. Address: Rua Gomes Carneiro, 1, Department of Economics, Campus Anglo, Universidade Federal de Pelotas, Centro, CEP 96010-610, Pelotas, RS, Brazil. Phone +55 (53) 3284-3865.}
\address[ufpel]{Department of Economics, Universidade Federal de Pelotas}
\address[ufsc]{Department of Economics and International Relations, Universidade Federal de Santa Catarina}

\begin{abstract}
This paper develops a doubly robust extension of local-projections difference-in-differences (LP-DiD) for staggered absorbing treatments. The resulting estimator, DRLPDID, preserves the LP-DiD local-stack ATT target and is consistent when either the local untreated-outcome regression or the local treatment-probability model is correctly specified. It also delivers influence-function-based inference for post-treatment summaries and multiplier-bootstrap bands for dynamic paths. In Monte Carlo designs with covariate-driven selection, DRLPDID matches regression-adjusted LP-DiD under outcome-model alignment and clearly outperforms the IPT-only variant under propensity-score misspecification. In the no-fault-divorce application, DRLPDID tracks robust staggered-adoption estimators and is less negative than unadjusted LP-DiD.
\end{abstract}


\end{frontmatter}


\section{Introduction}

Event-study designs are standard tools for tracing policy dynamics. In staggered-adoption settings with heterogeneous treatment effects, however, the usual two-way fixed-effects event-study regression need not recover a causal path because it can mix already-treated comparisons and non-transparent weights \citep{deChaisemartinDHaultfoeuille2020,GoodmanBacon2021,CallawaySantAnna2021,SunAbraham2021,BorusyakJaravelSpiess2024,Gardner2022,deChaisemartinDHaultfoeuille2023,Rothetal2023}.

LP-DiD estimates horizon-specific effects \citep{Jorda2005LocalProjections} on clean-control stacks that exclude unclean comparisons, thereby avoiding contamination from already-treated controls \citep{DubeGirardiJordaTaylor2025}. But each horizon-specific stack remains an observational comparison. When untreated outcome dynamics or treatment timing depend on covariates, the horizon-specific LP-DiD problem can be cast as a semiparametric Average Treatment Effect on the Treated (ATT) problem under conditional parallel trends \citep{DubeGirardiJordaTaylor2025}.
Although doubly robust estimators for staggered adoption already exist \citep{CallawaySantAnna2021,SantAnnaZhao2020}, they operate on the full panel rather than on horizon-specific clean-control stacks; a researcher committed to the LP-DiD framework, therefore, needs nuisance adjustment derived within that framework, with local propensity scores and outcome regressions that are stack- and horizon-specific by construction.

This article extends the LP-DiD framework by incorporating robustness to misspecification through a doubly robust approach in staggered adoption settings. It shows that the existing LP-DiD horizon problem can be enriched with Regression-Adjusted (RA), Inverse-Probability-Tilting (IPT), and Doubly Robust (DR) structures that admit influence-function-based inference and multiplier-bootstrap bands \citep{Hajek1971,GrahamPintoEgel2012,SantAnnaZhao2020,CallawaySantAnna2021}. The Monte Carlo compares DRLPDID with LP-DiD benchmarks and its RA and IPT counterparts across nuisance-specification scenarios designed to diagnose the value of double robustness. The application then shows how the resulting dynamic path compares with leading staggered-adoption estimators in practice. The appendix provides the general base-period formulation, proofs, and supplementary results.

\section{Local DRLPDID}

% Let $G_i$ be unit $i$'s first treatment date, with $G_i=\infty$ for never-treated units, and let treatment be binary and absorbing. LP-DiD compares post-entry outcomes with a pre-treatment benchmark
% \begin{equation}
% B_{it}^{(b)}=\sum_{\ell<t}\omega_{t\ell}^{(b)}Y_{i\ell},
% \end{equation}
% where the weights define the base-period rule. The paper uses the last pre-treatment benchmark, $B_{it}^{(-1)}=Y_{i,t-1}$, so
% \begin{equation}
% \Delta_{it}(h)=Y_{i,t+h}-Y_{i,t-1}.
% \end{equation}

LP-DiD compares post-entry outcomes with a pre-treatment benchmark constructed using a specific base-period rule, indexed by $b$. For a given entry date $t$, let $\mathcal{B}(t;b) \subset \{1, \dots, t-1\}$ denote the set of active pre-treatment periods and let $\omega_{tl}^{(b)}$ be weights summing to one. The generic benchmark operator is defined as:
\begin{equation}
B_{it}^{(b)} = \sum_{l \in \mathcal{B}(t;b)} \omega_{tl}^{(b)} Y_{il},
\end{equation}
This paper focuses on the standard last-pre-treatment benchmark (denoted by $b = -1$), which assigns a weight of one to period $t-1$ and zero otherwise, yielding $B_{it}^{(-1)} = Y_{i,t-1}$. The horizon-$h$ long difference is therefore:
\begin{equation}
\Delta_{it}(h) = Y_{i,t+h} - Y_{i,t-1}.
\end{equation}

Write the local treatment-entry indicator as
\begin{equation}
D_{it}=\ind\{G_i=t\}.
\end{equation}
Here $D_{it}$ denotes local treatment entry, not calendar-time treatment status.
For $h\geq 0$, the clean-control stack compares treated entries at $t$ with observations untreated at $t+h$. For $h<0$, it reuses the clean-control stack at $h=0$ and changes only the outcome transformation.

The horizon-specific local target is
\begin{equation}
\theta_h=\E\!\left[\Delta_{it}(h;1)-\Delta_{it}(h;0)\mid D_{it}=1,\ (i,t)\in\mathcal S_h\right].
\label{eq:theta_h}
\end{equation}
Let $X_{it}$ be predetermined covariates. For each $h$, assume no anticipation or overlap:
\begin{equation}
0<\Prb\!\left(D_{it}=1\mid X_{it},t,(i,t)\in\mathcal S_h\right)<1,
\end{equation}
and local conditional parallel trends,
\begin{equation}
\E\!\left[\Delta_{it}(h;0)\mid D_{it}=1,X_{it},t,(i,t)\in\mathcal S_h\right]
=
\E\!\left[\Delta_{it}(h;0)\mid D_{it}=0,X_{it},t,(i,t)\in\mathcal S_h\right].
\label{eq:cpt_h}
\end{equation}
Then each horizon is an ATT problem on its own LP-DiD stack.

Define the local nuisance functions
\begin{align}
\pi_h(X_{it},t)
&=\Prb\left(D_{it}=1\mid X_{it},t,(i,t)\in\mathcal S_h\right), \\
m_{0h}(X_{it},t)
&=\E\left[\Delta_{it}(h)\mid D_{it}=0,X_{it},t,(i,t)\in\mathcal S_h\right].
\end{align}
In the implemented estimators, both nuisance steps use the same horizon-specific linear basis $Q_{it,h}$, built from the same predetermined covariates and calendar-time controls; the difference is the fitting criterion, with inverse-probability tilting for the treatment-probability working model and least squares or weighted least squares for the untreated-outcome regression.

The local regression-adjusted and inverse-probability-weighting representations are
\begin{align}
\theta_h^{RA}
&=
\frac{\E\left[D_{it}\{\Delta_{it}(h)-m_{0h}(X_{it},t)\}\right]}
{\E[D_{it}]}, \\
\theta_h^{IPW}
&=
\frac{\E\left[D_{it}\Delta_{it}(h)\right]}{\E[D_{it}]}
-
\frac{\E\left[(1-D_{it})\omega_h(X_{it},t)\Delta_{it}(h)\right]}
{\E\left[(1-D_{it})\omega_h(X_{it},t)\right]},
\end{align}
where $\omega_h(X_{it},t)=\pi_h(X_{it},t)/(1-\pi_h(X_{it},t))$. Combining them yields
\begin{equation}
\theta_h^{DR}=
\frac{\E\left[D_{it}\{\Delta_{it}(h)-m_{0h}(X_{it},t)\}\right]}{\E[D_{it}]}
-
\frac{\E\left[(1-D_{it})\omega_h(X_{it},t)\{\Delta_{it}(h)-m_{0h}(X_{it},t)\}\right]}{\E\left[(1-D_{it})\omega_h(X_{it},t)\right]}.
\label{eq:dr_h}
\end{equation}
Under \eqref{eq:cpt_h}, \eqref{eq:dr_h} identifies $\theta_h$ if either $m_{0h}$ or $\pi_h$ is correctly specified. Here, IPW refers to the population weighting identity in \eqref{eq:dr_h}, whereas the implemented weighted estimator obtains $\pi_h$ by inverse-probability tilting (IPT). DRLPDID estimates \eqref{eq:dr_h} horizon by horizon using inverse-probability tilting for the local propensity score \citep{GrahamPintoEgel2012} and weighted least squares for the untreated-outcome regression, following the improved doubly robust logic of \citet{SantAnnaZhao2020} in an LP-DiD environment.

\subsection{Influence-function inference}

Let $\hat\Theta=(\hat\theta_h)_{h\in\mathcal H}$ collect the horizon-specific estimates. For each $h$, the improved estimator solves stacked local moments for the outcome-regression coefficients, the tilting parameters, and the treated and weighted-control means. Standard M-estimation then gives
\begin{equation}
\sqrt{N_C}(\hat\theta_h-\theta_h)=
\frac{1}{\sqrt{N_C}}\sum_{c=1}^{N_C}\IF_{c,h}+o_p(1),
\label{eq:if_h}
\end{equation}
where $N_C$ is the number of clusters. Equation \eqref{eq:if_h} yields cluster-robust standard errors for each horizon and for any linear contrast built from the horizon-specific effects.

For dynamic paths, the paper uses a multiplier bootstrap based on the stacked influence array. If $\hat\Phi_c=(\hat\IF_{c,h})_{h\in\mathcal H}$ and $\xi_c$ are i.i.d. multipliers with mean zero and variance one, bootstrap draws are
\begin{equation}
\hat\Theta^{\ast}=\hat\Theta+N_C^{-1/2}\sum_{c=1}^{N_C}\xi_c\hat\Phi_c.
\end{equation}
The main text reports pointwise cluster-robust intervals in the Monte Carlo and multiplier-bootstrap bands in the empirical event-study figures. The appendix provides the full moment system and proofs.

\section{Monte Carlo}

The Monte Carlo uses a staggered-adoption panel with covariate-driven treatment timing, heterogeneous dynamic treatment effects, and a never-treated group. The complete data-generating process is reported in the appendix. We compare five ATT-oriented estimators that share the same treated-entry-weighted target: LPDID-RW, LPDID-RW + X, LPDID-RA, DRLPDID-IPT, and DRLPDID. The designs vary regarding which nuisance component is aligned with the DGP: both nuisance components aligned, outcome-regression misspecified, propensity-score misspecified, and both nuisance components misspecified. Table~\ref{tab:sim_main} reports bias, RMSE, and pointwise 95\% coverage for the average post-treatment effect over 500 replications at $N=500$; the corresponding $N=250$ results are reported in Appendix Table~\ref{tab:sim_main_n250}.

\begin{table}[t!]
\centering
\caption{Monte Carlo: average post-treatment effect ($N=500$)}
\label{tab:sim_main}
\footnotesize
\begin{tabular}{llccc}
\toprule
Scenario & Estimator & Bias & RMSE & Coverage \\
\midrule
\multirow{5}{*}{\shortstack[l]{Both nuisances\\aligned}}
  & LPDID-RW      & -2.505 & 2.735 & 0.430 \\
  & LPDID-RW + X  &  0.064 & 0.104 & 0.866 \\
  & LPDID-RA      &  0.004 & 0.079 & 0.944 \\
  & DRLPDID-IPT   & -0.208 & 0.229 & 0.502 \\
  & DRLPDID       &  0.004 & 0.083 & 0.942 \\
\midrule
\multirow{5}{*}{OR misspecified}
  & LPDID-RW      & -3.561 & 3.747 & 0.100 \\
  & LPDID-RW + X  & -0.073 & 0.587 & 0.948 \\
  & LPDID-RA      & -0.108 & 0.577 & 0.958 \\
  & DRLPDID-IPT   & -0.308 & 0.634 & 0.904 \\
  & DRLPDID       &  0.011 & 0.634 & 0.924 \\
\midrule
\multirow{5}{*}{PS misspecified}
  & LPDID-RW      & -1.786 & 2.104 & 0.634 \\
  & LPDID-RW + X  &  0.061 & 0.102 & 0.872 \\
  & LPDID-RA      & -0.004 & 0.078 & 0.960 \\
  & DRLPDID-IPT   & -0.157 & 0.188 & 0.690 \\
  & DRLPDID       & -0.003 & 0.082 & 0.964 \\
\midrule
\multirow{5}{*}{\shortstack[l]{Both nuisances\\misspecified}}
  & LPDID-RW      & -3.543 & 3.712 & 0.098 \\
  & LPDID-RW + X  & -0.159 & 0.606 & 0.944 \\
  & LPDID-RA      & -0.214 & 0.614 & 0.938 \\
  & DRLPDID-IPT   & -0.382 & 0.682 & 0.896 \\
  & DRLPDID       & -0.068 & 0.663 & 0.916 \\
\bottomrule
\end{tabular}
\vspace{0.25em}

\parbox{0.92\linewidth}{\footnotesize Notes: All entries refer to the average post-treatment effect. Results are based on 500 Monte Carlo replications, with pointwise cluster-robust confidence intervals. The corresponding $N=250$ panel is reported in Appendix Table~\ref{tab:sim_main_n250}.}
\end{table}

Three patterns stand out. First, LPDID-RW is consistently unstable across scenarios once local selection and untreated trends depend on covariates. Second, when the outcome-regression component is aligned with the DGP, LPDID-RA remains highly competitive, and DRLPDID stays very close to it, rather than uniformly dominating it. Third, propensity-score misspecification is the setting in which the doubly robust construction is most informative: DRLPDID stays close to LPDID-RA, whereas DRLPDID-IPT deteriorates markedly. The low coverage of DRLPDID-IPT even in the aligned design appears to reflect bias rather than underestimated uncertainty: supplementary diagnostics show that its cluster-robust standard errors track the empirical dispersion closely, but the pure IPT step remains centered below the treated-entry-weighted target. This is consistent with the fact that covariate-balancing odds weights do not by themselves recover the cohort-size treated-entry weighting embedded in $\theta_h$, whereas the regression adjustment in DRLPDID corrects that discrepancy. Under joint misspecification, DRLPDID remains among the best scalar estimators, especially at $N=500$, but the Monte Carlo does not support a claim of uniform dominance in every design.

\section{Empirical application}

The application uses the no-fault-divorce panel from \citet{StevensonWolfers2006}, as revisited by \citet{GoodmanBacon2021}. The outcome, \texttt{asmrs}, is the state-level female suicide rate; treatment is the enactment of a no-fault divorce law. The sample covers the 48 continental states plus Washington, D.C., from 1964 to 1996. The baseline adjusted specification includes \texttt{asmrh}, the homicide mortality rate.\footnote{The original application also considers \texttt{pcinc} and \texttt{cases}. We omit them from the main specification because they may respond to the legal environment. The appendix reports a richer specification.}

Table~\ref{tab:app_att} reports average post-treatment effects under the parsimonious specification. DRLPDID yields $-5.69$, very close to Borusyak--Jaravel--Spiess and Gardner (both $-5.83$), less negative than Callaway--Sant'Anna ($-6.81$), and materially less negative than the two local LP-DiD comparators LPDID-RA ($-6.31$) and especially LPDID-RW + X ($-9.30$). Sun--Abraham remains clearly more negative at $-8.03$. The application therefore mirrors the Monte Carlo: covariate adjustment moves local LP-DiD toward the robust staggered-adoption comparators, and the doubly robust version aligns most closely with the imputation-based estimators Borusyak--Jaravel--Spiess and Gardner.

\begin{table}[t!]
\centering
\caption{Average post-treatment effects}
\label{tab:app_att}
\footnotesize
\begin{tabular}{lccc}
\toprule
Estimator & ATT & Std. Err. & 95\% CI \\
\midrule
DRLPDID                   &  -5.692 & 3.659 & [-12.864,\  1.479] \\
LPDID-RW + X              &  -9.304 & 3.612 & [-16.382,\ -2.225] \\
LPDID-RA                  &  -6.307 & 3.476 & [-13.120,\  0.505] \\
Callaway--Sant'Anna       &  -6.810 & 3.372 & [-13.003,\ -0.093] \\
Borusyak--Jaravel--Spiess &  -5.830 & 2.809 & [-11.322,\ -0.497] \\
Sun--Abraham              &  -8.029 & 3.326 & [-14.482,\ -1.354] \\
Gardner                   &  -5.830 & 3.025 & [-12.231,\ -0.287] \\
\bottomrule
\end{tabular}
\vspace{0.25em}

\parbox{0.92\linewidth}{\footnotesize Notes: LPDID-RW + X reports the covariate-adjusted local LP-DiD comparator from the Dube-family implementation. LPDID-RA reports the regression-adjusted local estimator from the semiparametric implementation. External staggered-adoption comparators are estimated with the Python \texttt{diff-diff} package \citep{diffdiff2026}.}
\end{table}

\begin{figure}[!htbp]
    \centering
    \includegraphics[width=0.95\textwidth]{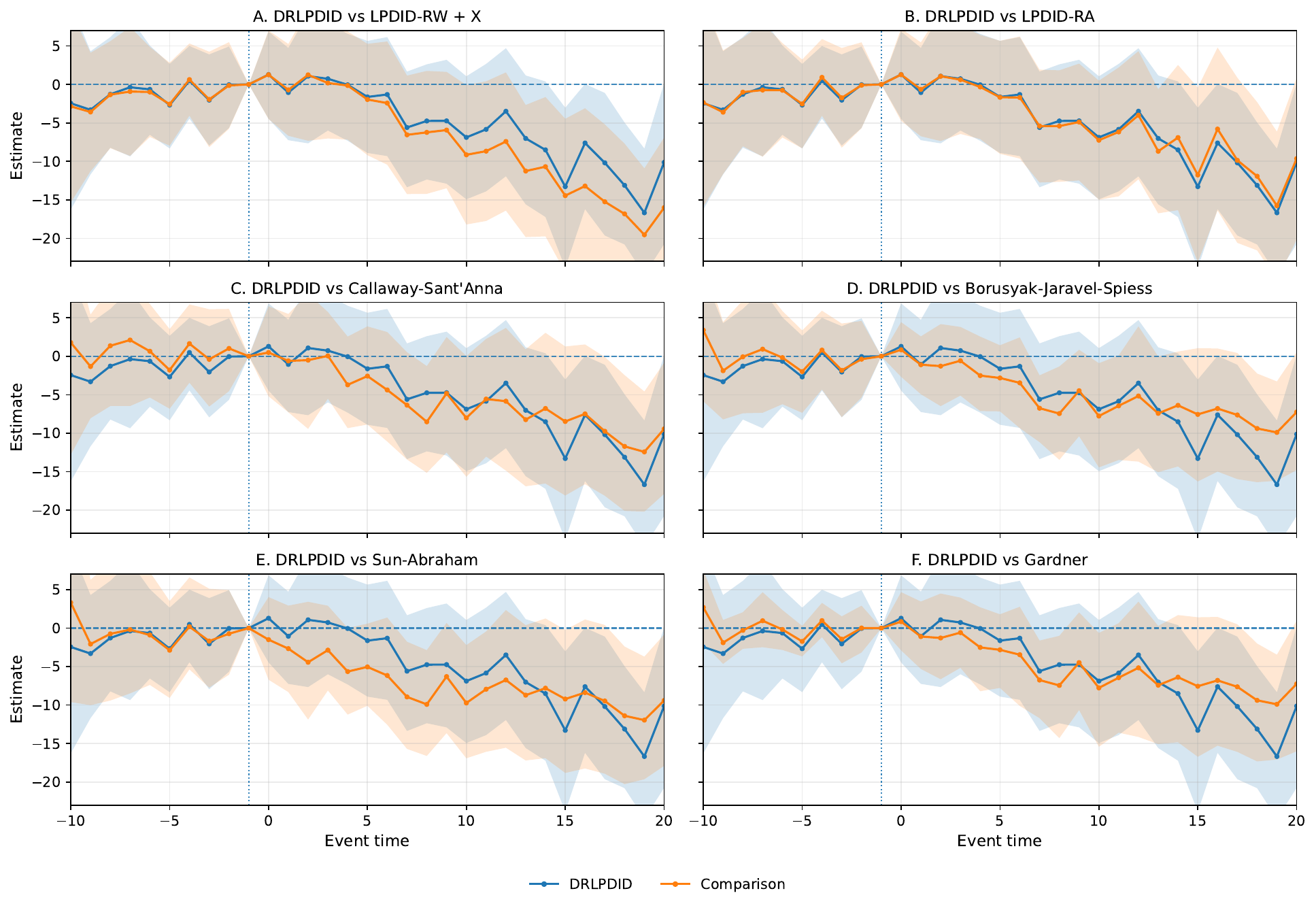}
    \caption{Dynamic event-study comparisons in the no-fault-divorce application}
    \label{fig:application_grid}
    \vspace{0.25em}

    \parbox{0.92\linewidth}{\footnotesize Notes: The five panels compare DRLPDID with LPDID-RW + X, Callaway--Sant'Anna, Borusyak--Jaravel--Spiess, Sun--Abraham, and Gardner under the same empirical specification. Shaded areas denote the reported confidence intervals from each estimator.}
\end{figure}

Figure~\ref{fig:application_grid} shows the dynamic paths. DRLPDID remains close to LPDID-RA and also tracks Borusyak--Jaravel--Spiess and Gardner more closely than the more negative LPDID-RW + X and Sun--Abraham paths over most post-treatment horizons. Callaway--Sant'Anna is also broadly in the same range, though somewhat more negative in the scalar summary.

\section{Conclusion}

DRLPDID extends LP-DiD with horizon-specific regression adjustment, inverse-probability tilting, and doubly robust moments without changing the local-stack target. Under the last pre-treatment benchmark used here, that target has the usual horizon-specific ATT interpretation. In the Monte Carlo, DRLPDID is consistently competitive and is especially well behaved under propensity-score misspecification, while LPDID-RA remains highly competitive when the outcome-regression component is well aligned with the DGP. In the no-fault-divorce application, DRLPDID delivers dynamic effects close to LPDID-RA and to the main robust staggered-adoption estimators, while remaining materially less negative than the IPW-only local path and the unadjusted LPDID-RW benchmark.

The practical implication is simple. When covariates plausibly shape local selection and untreated dynamics, unadjusted LP-DiD should not be the default specification. Extensions to continuous treatments, alternative base-period constructions, and stabilized treated controls are left for future work.

\section*{Funding}
This work was supported by the Conselho Nacional de Desenvolvimento Científico e Tecnológico (CNPq), grants 301475/2025-3 and 310327/2022-9.

\section*{Declaration of competing interests}
The authors declare that they have no known competing financial interests or personal relationships that could have appeared to influence the work reported in this paper.

\section*{Data availability}
The data and simulation materials underlying this article are available from the corresponding author upon reasonable request by email at \href{mailto:daniel.uhr@gmail.com}{daniel.uhr@gmail.com}.

\section*{Declaration of generative AI and AI-assisted technologies in the manuscript preparation process}
During the preparation of this work, the authors used ChatGPT and Claude Code for English-language editing and structural consistency checks. After using these tools, the authors reviewed and edited the manuscript as needed and take full responsibility for the content of the published article.


\begin{thebibliography}{99}

\bibitem[Borusyak et al.(2024)]{BorusyakJaravelSpiess2024}
Borusyak, K., X. Jaravel, J. Spiess, 2024.
Revisiting event-study designs: Robust and efficient estimation.
Review of Economic Studies 91, 3253--3285.

\bibitem[Callaway and Sant'Anna(2021)]{CallawaySantAnna2021}
Callaway, B., P. H. C. Sant'Anna, 2021.
Difference-in-differences with multiple time periods.
Journal of Econometrics 225, 200--230.

\bibitem[de Chaisemartin and D'Haultf{\oe}uille(2020)]{deChaisemartinDHaultfoeuille2020}
de Chaisemartin, C., X. D'Haultf{\oe}uille, 2020.
Two-way fixed effects estimators with heterogeneous treatment effects.
American Economic Review 110, 2964--2996.

\bibitem[de Chaisemartin and D'Haultf{\oe}uille(2023)]{deChaisemartinDHaultfoeuille2023}
de Chaisemartin, C., X. D'Haultf{\oe}uille, 2023.
Two-way fixed effects and difference-in-differences with heterogeneous treatment effects: A survey.
The Econometrics Journal 26, C1--C30.

\bibitem[diff-diff contributors(2026)]{diffdiff2026}
diff-diff contributors, 2026.
diff-diff: Difference-in-differences in Python.
Python package documentation. \url{https://diff-diff.readthedocs.io/}.

\bibitem[Dube et al.(2025)]{DubeGirardiJordaTaylor2025}
Dube, A., D. Girardi, \`O. Jord\`a, A. M. Taylor, 2025.
A local projections approach to difference-in-differences.
Journal of Applied Econometrics 40, 741--758.

\bibitem[Gardner(2022)]{Gardner2022}
Gardner, J., 2022.
Two-stage differences in differences.
arXiv preprint arXiv:2207.05943.

\bibitem[Goodman-Bacon(2021)]{GoodmanBacon2021}
Goodman-Bacon, A., 2021.
Difference-in-differences with variation in treatment timing.
Journal of Econometrics 225, 254--277.

\bibitem[Graham et al.(2012)]{GrahamPintoEgel2012}
Graham, B. S., C. C. Pinto, D. Egel, 2012.
Inverse probability tilting for moment condition models with missing data.
Review of Economic Studies 79, 1053--1079.

\bibitem[H{\'a}jek(1971)]{Hajek1971}
H{\'a}jek, J., 1971.
Discussion of ``An Essay on the Logical Foundations of Survey Sampling, Part I'', by D. Basu.
In: Godambe, V.P., Sprott, D.A. (Eds.),
Foundations of Statistical Inference. Holt, Rinehart and Winston,
Toronto, p. 326.

\bibitem[Jorda(2005)]{Jorda2005LocalProjections}
Jord{\'a}, \`O., 2005.
Estimation and inference of impulse responses by local projections.
American Economic Review 95, 161--182.

\bibitem[Roth et al.(2023)]{Rothetal2023}
Roth, J., M. Sant'Anna, A. Bilinski, J. Poe, 2023.
What’s trending in difference-in-differences? A synthesis of the recent econometrics literature.
Journal of Econometrics 235, 2218--2244.

\bibitem[Sant'Anna and Zhao(2020)]{SantAnnaZhao2020}
Sant'Anna, P. H. C., J. Zhao, 2020.
Doubly robust difference-in-differences estimators.
Journal of Econometrics 219, 101--122.

\bibitem[Stevenson and Wolfers(2006)]{StevensonWolfers2006}
Stevenson, B., J. Wolfers, 2006.
Bargaining in the shadow of the law: Divorce laws and family distress.
Quarterly Journal of Economics 121, 267--288.

\bibitem[Sun and Abraham(2021)]{SunAbraham2021}
Sun, L., S. Abraham, 2021.
Estimating dynamic treatment effects in event studies with heterogeneous treatment effects.
Journal of Econometrics 225, 175--199.

\end{thebibliography}
\end{document}

\begin{frontmatter}

\title{Online Appendix for \emph{Doubly Robust Local Projections Difference-in-Differences}}

\author[ufpel]{Daniel de Abreu Pereira Uhr}
\ead{daniel.uhr@gmail.com}
\author[ufsc]{Guilherme Valle Moura}
\ead{gvallemoura@gmail.com}
\address[ufpel]{Department of Economics, Universidade Federal de Pelotas}
\address[ufsc]{Department of Economics and International Relations, Universidade Federal de Santa Catarina}

\end{frontmatter}

This online appendix provides the theoretical and empirical details omitted from the main text. It is organized into seven sections. 
The first three sections introduce the general LP-DiD notation and clean-control stack construction, the mapping from the local target to cohort-time treatment effects, and the two aggregation rules used in the paper: the benchmark variance-weighted LP-DiD aggregation and the treated-entry-weighted target underlying the RW, RA, IPT, and DR estimators. 
The next two sections present the regression-adjustment, inverse-probability-weighting representation with an inverse-probability-tilting first step, and doubly robust representations, together with the asymptotic linear representation used for cluster-robust inference and multiplier-bootstrap bands. 
The final two sections report the Monte Carlo design and additional results, including the data-generating process actually used in the final simulations, a supplementary sensitivity exercise to violations of local conditional parallel trends, and the richer-covariate specification used in the no-fault-divorce application.

\appendix
\renewcommand{\thesection}{Appendix \Alph{section}}
\renewcommand{\thesubsection}{\Alph{section}.\arabic{subsection}}
\renewcommand{\theequation}{\Alph{section}.\arabic{equation}}
\renewcommand{\thetable}{\Alph{section}.\arabic{table}}
\renewcommand{\thefigure}{\Alph{section}.\arabic{figure}}
\renewcommand{\theassumption}{\Alph{section}.\arabic{assumption}}
\renewcommand{\thedefinition}{\Alph{section}.\arabic{assumption}}
\renewcommand{\theproposition}{\Alph{section}.\arabic{assumption}}
\renewcommand{\theremark}{\Alph{section}.\arabic{assumption}}

\section{General LP-DiD notation and clean-control stacks}

Let $i=1,\dots,N$ index units and $t=1,\dots,T$ periods. Treatment is binary and absorbing, with the first treatment date $G_i\in\{1,\dots,T,\infty\}$ and $G_i=\infty$ for never-treated units. Define the treatment-entry indicator
\begin{equation}
D_{it}=\ind\{G_i=t\}.
\end{equation}

\begin{definition}[Base-period operator]
For a generic rule $b$, let $\BB(t;b)\subset\{1,\dots,t-1\}$ denote the set of pre-treatment periods used to construct the comparison base for entry date $t$, and let $\{\omega_{t\ell}^{(b)}:\ell\in\BB(t;b)\}$ be weights satisfying
\begin{equation}
\sum_{\ell\in\BB(t;b)}\omega_{t\ell}^{(b)}=1.
\end{equation}
The associated base-period operator is
\begin{equation}
B_{it}^{(b)}=\sum_{\ell\in\BB(t;b)}\omega_{t\ell}^{(b)}Y_{i\ell}.
\end{equation}
The corresponding long difference at horizon $h\in\HH$ is
\begin{equation}
\Delta_{it}^{(b)}(h)=Y_{i,t+h}-B_{it}^{(b)}.
\end{equation}
\end{definition}

The main text uses the last-pre-treatment rule $b=-1$, for which $\BB(t;-1)=\{t-1\}$, $\omega_{t,t-1}^{(-1)}=1$, and
\begin{equation}
B_{it}^{(-1)}=Y_{i,t-1},
\qquad
\Delta_{it}^{(-1)}(h)=Y_{i,t+h}-Y_{i,t-1}.
\end{equation}
Other admissible LP-DiD implementations may use a different pre-treatment operator, such as an average over several pre-treatment periods, provided that the comparison base remains supported by untreated periods relative to the entry date.

For a given entry date $t$ and horizon $h$, define the comparison window
\begin{equation}
\WW_h(t;b)=\{t+h\}\cup \BB(t;b).
\end{equation}
Because treatment is absorbing, a control unit is clean for $(t,h,b)$ if it is untreated throughout the comparison window. Hence, the clean-control set is
\begin{equation}
\CC_h(t;b)=\{i: T_{is}=0\ \text{for all } s\in \WW_h(t;b)\},
\end{equation}
where $T_{is}=\ind\{s\geq G_i,\ G_i<\infty\}$ denotes calendar-time treatment status. The local LP-DiD stack is
\begin{equation}
\SSS_h(t;b)=\{i:D_{it}=1\}\cup \CC_h(t;b).
\end{equation}
Pooling over admissible entry dates gives
\begin{equation}
\SSS_h^{(b)}=\bigcup_{t:\,t+h\in\{1,\dots,T\}}\{(i,t):i\in\SSS_h(t;b)\}.
\end{equation}
Throughout the appendix, $D_{it}$ denotes local treatment entry, while $T_{is}$ denotes calendar-time treatment status.
For $h\geq 0$ and $b=-1$, clean controls are exactly the units untreated at $t+h$. For $h<0$, the clean-control condition collapses to being untreated at the entry date $t$, because every period in the comparison window is weakly before $t$.

\begin{definition}[Horizon-specific local target]
For a given base rule $b$, the LP-DiD horizon-$h$ target is
\begin{equation}
\theta_h^{(b)}=
\E\left[\Delta_{it}^{(b)}(h;1)-\Delta_{it}^{(b)}(h;0)\mid D_{it}=1,\ (i,t)\in\SSS_h^{(b)}\right],
\label{eq:def_thetah}
\end{equation}
where $\Delta_{it}^{(b)}(h;d)$ denotes the long difference constructed from the potential-outcome path under treatment state $d\in\{0,1\}$.
\end{definition}

\section{From LP-DiD stacks to cohort-time treatment effects}

To make the cohort structure explicit, fix a cohort $g<\infty$ and write
\begin{equation}
\theta_{g,h}^{(b)}=
\E\left[\Delta_{ig}^{(b)}(h;1)-\Delta_{ig}^{(b)}(h;0)\mid G_i=g\right],
\qquad h\geq 0.
\end{equation}
Also, define the usual cohort-time ATT
\begin{equation}
\ATT(g,g+h)=\E\left[Y_{i,g+h}(1)-Y_{i,g+h}(0)\mid G_i=g\right].
\end{equation}

\begin{assumption}[No anticipation relative to the base operator]
For every treated cohort $g$ and every pre-treatment period $\ell<g$, treatment does not affect outcomes before entry:
\begin{equation}
Y_{i\ell}(1)=Y_{i\ell}(0)
\qquad \text{a.s. conditional on } G_i=g.
\end{equation}
\end{assumption}

\begin{proposition}[Mapping from the local target to $\ATT(g,g+h)$]
\label{prop:mapping}
Suppose the base-period operator uses only pre-treatment periods and the weights sum to one. Then, for every $g<\infty$ and every post-treatment horizon $h\geq 0$,
\begin{equation}
\theta_{g,h}^{(b)}=\ATT(g,g+h).
\end{equation}
Hence any LP-DiD target of the form \eqref{eq:def_thetah} can be written as a weighted average of cohort-time treatment effects. In particular, under the last-pre-treatment rule $b=-1$ used in the paper, the horizon-specific local target coincides with the usual horizon-specific ATT target.
\end{proposition}

\begin{proof}
By definition,
\begin{align}
\theta_{g,h}^{(b)}
&=\E\left[Y_{i,g+h}(1)-\sum_{\ell\in\BB(g;b)}\omega_{g\ell}^{(b)}Y_{i\ell}(1)
      -Y_{i,g+h}(0)+\sum_{\ell\in\BB(g;b)}\omega_{g\ell}^{(b)}Y_{i\ell}(0)
      \mid G_i=g\right].
\end{align}
Under no anticipation, $Y_{i\ell}(1)=Y_{i\ell}(0)$ for every $\ell<g$. Since $\BB(g;b)\subset\{1,\dots,g-1\}$, the baseline terms cancel exactly. Therefore
\begin{equation}
\theta_{g,h}^{(b)}=\E\left[Y_{i,g+h}(1)-Y_{i,g+h}(0)\mid G_i=g\right]=\ATT(g,g+h).
\end{equation}
The final statement follows by specializing to $b=-1$.
\end{proof}

\begin{remark}
What changes with the choice of $b$ is not the post-treatment cohort-time estimand itself, but the normalization of the long-difference outcome and the finite-sample implementation of the local stack. The paper uses the last-pre-treatment benchmark because it is the empirical default in the LP-DiD implementation studied here and keeps the notation concise.
\end{remark}

\section{Benchmark LP-DiD aggregation and the ATT-oriented target}

This section complements the main text by making the aggregation step explicit. Fix a post-treatment horizon $h\geq 0$ and specialize to $b=-1$. For each cohort $g<\infty$, define the cohort-specific clean-control cell
\begin{equation}
\mathcal C_{g,h}=\{i:G_i=g\}\cup\{i:G_i>g+h\ \text{or } G_i=\infty\}.
\end{equation}
Let $N_g$ denote the number of treated entrants in cohort $g$, let $N_{g,h}^{0}$ denote the number of clean controls in that cell, and write $N_{g,h}=N_g+N_{g,h}^{0}$ and $n_{g,h}=N_g/N_{g,h}$.

Within cell $(g,h)$, the clean DID contrast is
\begin{equation}
\delta_{g,h}=\E\big[Y_{i,g+h}-Y_{i,g-1}\mid G_i=g\big]
-\E\big[Y_{i,g+h}-Y_{i,g-1}\mid i\in\mathcal C_{g,h},\ D_{ig}=0\big].
\end{equation}
Under no anticipation and clean-control parallel trends, $\delta_{g,h}=\ATT(g,g+h)$.

Consider the pooled LP-DiD regression with cohort-horizon intercepts,
\begin{equation}
Y_{i,g+h}-Y_{i,g-1}=\alpha_h D_{ig}+\lambda_{gh}+u_{igh},
\label{eq:benchmark_reg}
\end{equation}
on the union of the cohort-horizon clean cells.

\begin{proposition}[Benchmark LP-DiD weighting formula]
\label{prop:benchmark_weights}
The population coefficient on $D_{ig}$ in \eqref{eq:benchmark_reg} is
\begin{equation}
\alpha_h^{VW}=\sum_{g:G_g<\infty}\omega_{g,h}^{VW}\,\ATT(g,g+h),
\end{equation}
with weights
\begin{equation}
\omega_{g,h}^{VW}=\frac{N_{g,h}n_{g,h}(1-n_{g,h})}{\sum_{g^{\prime}:G_{g^{\prime}}<\infty}N_{g',h}n_{g',h}(1-n_{g',h})}.
\end{equation}
\end{proposition}

\begin{proof}
By Frisch--Waugh--Lovell, the coefficient on $D_{ig}$ is obtained by residualizing $D_{ig}$ with respect to the cohort-horizon intercepts. Inside cell $(g,h)$, the residualized treatment indicator is
\begin{equation}
\widetilde D_{ig}=D_{ig}-\bar D_{g,h}=D_{ig}-n_{g,h}.
\end{equation}
Hence, the cell-specific denominator contribution is
\begin{equation}
\sum_{i\in\mathcal C_{g,h}}\widetilde D_{ig}^{2}=N_{g,h}n_{g,h}(1-n_{g,h}),
\end{equation}
while the numerator contribution is the same variance term multiplied by the cell DID contrast $\delta_{g,h}$. Summing across cohorts yields
\begin{equation}
\alpha_h^{VW}=\frac{\sum_g N_{g,h}n_{g,h}(1-n_{g,h})\delta_{g,h}}{\sum_g N_{g,h}n_{g,h}(1-n_{g,h})}.
\end{equation}
Substituting $\delta_{g,h}=\ATT(g,g+h)$ proves the result.
\end{proof}

\begin{proposition}[RW re-weighting recovers the treated-entry-weighted ATT]
\label{prop:rw_weights}
Suppose \eqref{eq:benchmark_reg} is estimated by weighted least squares, assigning a common positive weight $\lambda_{g,h}$ to all observations in cell $\mathcal C_{g,h}$. Then the population coefficient on $D_{ig}$ is
\begin{equation}
\alpha_h(\lambda)=\sum_{g:G_g<\infty}\omega_{g,h}(\lambda)\,\ATT(g,g+h),
\end{equation}
with
\begin{equation}
\omega_{g,h}(\lambda)=
\frac{\lambda_{g,h}N_{g,h}n_{g,h}(1-n_{g,h})}
{\sum_{g':G_g'<\infty}\lambda_{g',h}N_{g',h}n_{g',h}(1-n_{g',h})}.
\end{equation}
In particular, if the cell weights satisfy
\begin{equation}
\lambda_{g,h}^{RW}\propto
\left(\frac{\omega_{g,h}^{VW}}{N_g}\right)^{-1},
\label{eq:rw_choice}
\end{equation}
equivalently,
\begin{equation}
\lambda_{g,h}^{RW}\propto
\left(\frac{N_{g,h}n_{g,h}(1-n_{g,h})}{N_g}\right)^{-1}
=\frac{1}{1-n_{g,h}}
=\frac{N_{g,h}}{N_{g,h}-N_g},
\label{eq:rw_choice_simplified}
\end{equation}
then
\begin{equation}
\omega_{g,h}^{RW}=
\frac{N_g}{\sum_{g^{\prime}:G_{g^{\prime}}<\infty}N_{g'}}
\end{equation}
and therefore
\begin{equation}
\alpha_h^{RW}
=
\frac{\sum_{g:G_g<\infty}N_g\,\ATT(g,g+h)}
{\sum_{g:G_g<\infty}N_g}
=
\theta_h.
\end{equation}
\end{proposition}

\begin{proof}
If a constant weight $\lambda_{g,h}$ is assigned to all observations in cell $\mathcal C_{g,h}$, Frisch--Waugh--Lovell implies that the residualized treatment indicator remains
\begin{equation}
\widetilde D_{ig}=D_{ig}-n_{g,h}
\qquad \text{for } i\in\mathcal C_{g,h},
\end{equation}
because the weight does not vary within the cell. Hence, the weighted cell-specific denominator contribution is
\begin{equation}
\sum_{i\in\mathcal C_{g,h}}\lambda_{g,h}\widetilde D_{ig}^{2}
=
\lambda_{g,h}N_{g,h}n_{g,h}(1-n_{g,h}),
\end{equation}
and the weighted numerator contribution is the same term multiplied by the cell DID contrast $\delta_{g,h}$. Therefore
\begin{equation}
\alpha_h(\lambda)=
\frac{\sum_g \lambda_{g,h}N_{g,h}n_{g,h}(1-n_{g,h})\delta_{g,h}}
{\sum_g \lambda_{g,h}N_{g,h}n_{g,h}(1-n_{g,h})}.
\end{equation}
Substituting $\delta_{g,h}=\ATT(g,g+h)$ yields the first result.

Now, choose $\lambda_{g,h}=\lambda_{g,h}^{RW}$ as in \eqref{eq:rw_choice}. Since
\begin{equation}
\omega_{g,h}^{VW}\propto N_{g,h}n_{g,h}(1-n_{g,h}),
\end{equation}
we obtain
\begin{equation}
\lambda_{g,h}^{RW}N_{g,h}n_{g,h}(1-n_{g,h})
\propto
\left(\frac{N_{g,h}n_{g,h}(1-n_{g,h})}{N_g}\right)^{-1}
N_{g,h}n_{g,h}(1-n_{g,h})
=
N_g.
\end{equation}
Thus, each cell contributes proportionally to $N_g$, so the normalized aggregation weights become
\begin{equation}
\omega_{g,h}^{RW}
=
\frac{N_g}{\sum_{g^{\prime}:G_{g^{\prime}}<\infty}N_{g'}}.
\end{equation}
Substituting into the aggregation formula gives
\begin{equation}
\alpha_h^{RW}
=
\frac{\sum_{g:G_g<\infty}N_g\,\ATT(g,g+h)}
{\sum_{g:G_g<\infty}N_g}
=
\theta_h,
\end{equation}
which proves the claim.
\end{proof}

The benchmark LP-DiD coefficient is therefore a clean-comparison object, but it is not yet treated-entry weighted. Proposition \ref{prop:rw_weights} shows that the RW re-weighting of \cite{DubeGirardiJordaTaylor2025} removes the extra factor $(1-n_{g,h})$ embedded in the benchmark variance weights and converts the aggregation into direct weighting by cohort size. The resulting target is
\begin{equation}
\theta_h=\frac{\sum_{g:G_g<\infty}N_g\ATT(g,g+h)}{\sum_{g:G_g<\infty}N_g}.
\label{eq:att_target_h}
\end{equation}
This is the same aggregation target pursued by the local RA, IPT, and DR estimators studied in the main text.

\section{Local semiparametric identification}

For the remainder of the appendix, let $X_{it}$ denote predetermined covariates and specialize in the clean-control stack actually used in the paper. Identification is horizon specific.

\begin{assumption}[Local conditional parallel trends]
\label{ass:cpt}
For every reported horizon $h$,
\begin{equation}
\hspace{-0.2cm}\E\left[\Delta_{it}(h;0)\!\mid\!\! D_{it}\!=\!1,X_{it},t,(i,t)\!\in\!\SSS_h^{(-1)}\right]
\!=\!
\E\left[\Delta_{it}(h;0)\!\mid\!\! D_{it}\!=\!0,X_{it},t,(i,t)\in\SSS_h^{(-1)}\right].
\end{equation}
\end{assumption}

\begin{assumption}[Overlap]
\label{ass:overlap}
For every reported horizon $h$,
\begin{equation}
0<\pi_h(X_{it},t)<1
\qquad \text{a.s. on } \SSS_h^{(-1)},
\end{equation}
where
\begin{equation}
\pi_h(X_{it},t)=\Prb\left(D_{it}=1\mid X_{it},t,(i,t)\in\SSS_h^{(-1)}\right).
\end{equation}
\end{assumption}

Define also the untreated long-difference regression
\begin{equation}
m_{0h}(X_{it},t)=\E\left[\Delta_{it}(h)\mid D_{it}=0,X_{it},t,(i,t)\in\SSS_h^{(-1)}\right].
\end{equation}

\subsection{Regression adjustment}

\begin{proposition}[Local RA representation]
\label{prop:ra}
Under Assumptions \ref{ass:cpt} and \ref{ass:overlap},
\begin{equation}
\theta_h=
\frac{\E\left[D_{it}\{\Delta_{it}(h)-m_{0h}(X_{it},t)\}\right]}{\E[D_{it}]}.
\label{eq:ra_repr}
\end{equation}
\end{proposition}

\begin{proof}
By Assumption \ref{ass:cpt},
\begin{equation}
\E\left[\Delta_{it}(h;0)\mid D_{it}=1,X_{it},t,(i,t)\in\SSS_h^{(-1)}\right]=m_{0h}(X_{it},t).
\end{equation}
Therefore,
\begin{align}
\hspace{-0.8cm}\E\left[\Delta_{it}(h)\!-\!m_{0h}(X_{it},t)\!\mid\!\! D_{it}\!=\!1,(i,t)\!\in\!\SSS_h^{(-1)}\right]\!
&\!=\!\E\left[\Delta_{it}(h;1)\!-\!\Delta_{it}(h;0)\!\mid\!\! D_{it}\!=\!1,(i,t)\!\in\!\SSS_h^{(-1)}\right]
\\
&=\theta_h.
\end{align}
Multiplying by $\Prb(D_{it}=1)=\E[D_{it}]$ yields \eqref{eq:ra_repr}.
\end{proof}

\subsection{Inverse-probability weighting and inverse-probability tilting}

Define the local odds weight
\begin{equation}
\omega_h(X_{it},t)=\frac{\pi_h(X_{it},t)}{1-\pi_h(X_{it},t)}.
\end{equation}
The population IPW representation uses this weight, while the estimator in the paper obtains $\pi_h$ through inverse-probability tilting \citep{GrahamPintoEgel2012}. The distinction is between the weighting identity and the first-stage fitting criterion; the target moment is the same.

\begin{proposition}[Local IPW representation]
\label{prop:ipw}
Under Assumptions \ref{ass:cpt} and \ref{ass:overlap},
\begin{equation}
\theta_h=
\frac{\E\left[D_{it}\Delta_{it}(h)\right]}{\E[D_{it}]}
-
\frac{\E\left[(1-D_{it})\omega_h(X_{it},t)\Delta_{it}(h)\right]}{\E\left[(1-D_{it})\omega_h(X_{it},t)\right]}.
\label{eq:ipw_repr}
\end{equation}
\end{proposition}

\begin{proof}
For any integrable random variable $Z_{it}$ measurable with respect to $(X_{it},t)$ on the local stack,
\begin{align}
\E\left[(1-D_{it})\omega_h(X_{it},t)Z_{it}\mid X_{it},t\right]
&=\omega_h(X_{it},t)\{1-\pi_h(X_{it},t)\}Z_{it}
\\
&=\pi_h(X_{it},t)Z_{it}
\\
&=\E\left[D_{it}Z_{it}\mid X_{it},t\right].
\end{align}
Applying this identity to $Z_{it}=1$ gives
\begin{equation}
\E\left[(1-D_{it})\omega_h(X_{it},t)\right]=\E[D_{it}].
\label{eq:denominator_identity}
\end{equation}
Applying it to $Z_{it}=\Delta_{it}(h;0)$ yields
\begin{equation}
\E\left[(1-D_{it})\omega_h(X_{it},t)\Delta_{it}(h;0)\right]=\E\left[D_{it}\Delta_{it}(h;0)\right].
\end{equation}
Subtract the latter equality from $\E[D_{it}\Delta_{it}(h)]$ and use the fact that, on treated entries, $\Delta_{it}(h)=\Delta_{it}(h;1)$ to obtain
\begin{equation}
\E[D_{it}]\theta_h
=
\E\left[D_{it}\Delta_{it}(h)\right]-\E\left[(1-D_{it})\omega_h(X_{it},t)\Delta_{it}(h)\right].
\end{equation}
Dividing both sides by the common denominator gives \eqref{eq:ipw_repr}.
\end{proof}

\subsection{Local inverse-probability-tilting first step}

To make the first-stage fitting step explicit, let $Q_{it,h}$ denote the common horizon-$h$ linear basis used in both nuisance models. In the implementation used in the paper, $Q_{it,h}$ contains the same predetermined covariates and the same calendar-time controls in both first stages. The local propensity-score working model is
\begin{equation}
\pi_h(X_{it},t;\gamma_h)=\Lambda\!\left(Q_{it,h}'\gamma_h\right),
\,\,
\omega_h(X_{it},t;\gamma_h)=\frac{\pi_h(X_{it},t;\gamma_h)}{1-\pi_h(X_{it},t;\gamma_h)}
=\exp\!\left(Q_{it,h}'\gamma_h\right),
\label{eq:ipt_logit_local}
\end{equation}
and the untreated-outcome regression working model is
\begin{equation}
m_{0h}(X_{it},t;\beta_h)=Q_{it,h}'\beta_h.
\end{equation}
The IPT estimator chooses $\hat\gamma_h$ to satisfy the tilting first-order condition
\begin{equation}
\sum_{(i,t)\in\SSS_h^{(-1)}} Q_{it,h}
\left\{
D_{it}-(1-D_{it})\exp\!\left(Q_{it,h}'\hat\gamma_h\right)
\right\}=0.
\label{eq:ipt_moment_local}
\end{equation}
Equivalently, $\hat\gamma_h$ solves the sample tilting problem
\begin{equation}
\hat\gamma_h\in
\arg\max_{\gamma}
\sum_{(i,t)\in\SSS_h^{(-1)}}
\left[
D_{it} Q_{it,h}'\gamma
-
(1-D_{it})\exp\!\left(Q_{it,h}'\gamma\right)
\right].
\label{eq:ipt_objective_local}
\end{equation}
Given $\hat\gamma_h$, the improved untreated-outcome regression is fitted on the local controls by weighted least squares with the same linear basis and odds weights,
\begin{equation}
\hat\beta_h\in
\arg\min_{\beta}
\sum_{(i,t)\in\SSS_h^{(-1)}}
(1-D_{it})\hat\omega_h(X_{it},t)
\left[
\Delta_{it}(h)-Q_{it,h}'\beta
\right]^2,
\label{eq:wls_or_local}
\end{equation}
where $\hat\omega_h(X_{it},t)=\exp(Q_{it,h}'\hat\gamma_h)$. Thus, the improved local LP-DiD estimator uses the same covariate-time basis in the IPT and OR steps, differing only in the fitting criterion.

\subsection{Doubly robust representation}

\begin{proposition}[Local DR representation and double robustness]
\label{prop:dr}
Define
\begin{equation}
\theta_h^{DR}=
\frac{\E\left[D_{it}\{\Delta_{it}(h)-m_{0h}(X_{it},t)\}\right]}{\E[D_{it}]}
-
\frac{\E\left[(1-D_{it})\omega_h(X_{it},t)\{\Delta_{it}(h)-m_{0h}(X_{it},t)\}\right]}{\E\left[(1-D_{it})\omega_h(X_{it},t)\right]}.
\label{eq:dr_repr}
\end{equation}
Then \eqref{eq:dr_repr} identifies $\theta_h$ if either the untreated-outcome regression $m_{0h}$ is correctly specified or the local propensity score $\pi_h$ is correctly specified.
\end{proposition}

\begin{proof}
If $m_{0h}$ is correctly specified, then by Assumption \ref{ass:cpt},
\begin{equation}
\E\left[\Delta_{it}(h)-m_{0h}(X_{it},t)\mid D_{it}=0,X_{it},t,(i,t)\in\SSS_h^{(-1)}\right]=0.
\end{equation}
Consequently, the weighted control residual in \eqref{eq:dr_repr} has mean zero for any measurable weighting function with finite expectation, including a misspecified odds weight. The first term therefore reduces to the RA representation and identifies $\theta_h$.

If instead $\pi_h$ is correctly specified while $m_{0h}$ may be misspecified, write $m_h(X_{it},t)$ for the possibly misspecified regression entering \eqref{eq:dr_repr}. Since $\pi_h$ is correctly specified, the odds weight satisfies $\omega_h(X_{it},t)\{1-\pi_h(X_{it},t)\}=\pi_h(X_{it},t)$ exactly. By the same weighting identity used in the proof of Proposition \ref{prop:ipw},
\begin{equation}
\E\left[(1-D_{it})\omega_h(X_{it},t)m_h(X_{it},t)\right]
=
\E\left[D_{it}m_h(X_{it},t)\right].
\label{eq:dr_cancel_local}
\end{equation}
Using also \eqref{eq:denominator_identity}, the DR moment can be written as
\begin{align}
\hspace{-2cm}\theta_h^{DR}
&=
\frac{\E[D_{it}\Delta_{it}(h)]-\E[D_{it}m_h(X_{it},t)]}{\E[D_{it}]}
-
\frac{\E[(1-D_{it})\omega_h(X_{it},t)\Delta_{it}(h)]-\E[(1-D_{it})\omega_h(X_{it},t)m_h(X_{it},t)]}{\E[D_{it}]}
\\
\hspace{-2cm}&=
\frac{\E[D_{it}\Delta_{it}(h)]-\E[(1-D_{it})\omega_h(X_{it},t)\Delta_{it}(h)]}{\E[D_{it}]},
\end{align}
where the second line follows from \eqref{eq:dr_cancel_local}. Thus, the regression term cancels exactly between the treated and weighted-control components, leaving the correctly weighted IPW representation. Therefore, the DR moment continues to identify $\theta_h$.
\end{proof}

\begin{remark}[The improved DR estimator]
The improved estimator used in the paper does not alter the target moment. It alters only the nuisance-estimation equations. Specifically, the local propensity score is estimated by inverse-probability tilting rather than by logit maximum likelihood, and the untreated-outcome regression is fit by weighted least squares with odds weights as in \eqref{eq:wls_or_local}, mirroring the improved doubly robust construction of \citet{SantAnnaZhao2020} in a horizon-specific LP-DiD environment. The estimand remains \eqref{eq:dr_repr}; the gain concerns first-order robustness of inference.
\end{remark}

\section{Influence-function-based inference}

For each horizon $h$, let $\eta_h$ collect the nuisance and target parameters entering the local improved-DR estimation problem. In the implementation used in the paper,
\begin{equation}
\eta_h=(\beta_h',\gamma_h',\mu_{1h},\mu_{0h})',
\end{equation}
where $m_{0h}(X_{it},t)=Q_{it,h}'\beta_h$, $\pi_h(X_{it},t)=\Lambda(Q_{it,h}'\gamma_h)$, and $\theta_h=\mu_{1h}-\mu_{0h}$. Let the cluster-level moment vector be $\Psi_{c,h}(\eta_h)$, where $c$ indexes the sampling clusters. In the implementation studied here, $\Psi_{c,h}(\eta_h)$ stacks the weighted-least-squares normal equation for $\beta_h$, the IPT score equation in \eqref{eq:ipt_moment_local} for $\gamma_h$, and the moment conditions defining the treated and weighted-control means $\mu_{1h}$ and $\mu_{0h}$.

\begin{proposition}[Cluster-level asymptotic linear representation]
\label{prop:if}
Suppose the estimator solves
\begin{equation}
\frac{1}{N_C}\sum_{c=1}^{N_C}\Psi_{c,h}(\hat\eta_h)=0,
\end{equation}
and assume standard smoothness, identification, and finite-moment conditions for cluster-level M-estimation, with a nonsingular Jacobian
\begin{equation}
A_h=\E\left[\frac{\partial \Psi_{c,h}(\eta_h)}{\partial \eta_h'}\right]\Bigg|_{\eta_h=\eta_h^0}.
\end{equation}
Then
\begin{equation}
\sqrt{N_C}(\hat\theta_h-\theta_h)
=
\frac{1}{\sqrt{N_C}}\sum_{c=1}^{N_C}\IF_{c,h}+o_p(1),
\label{eq:if_linear}
\end{equation}
where
\begin{equation}
\IF_{c,h}=-\nabla_{\eta}g(\eta_h^0)'A_h^{-1}\Psi_{c,h}(\eta_h^0),
\qquad g(\eta_h)=\mu_{1h}-\mu_{0h}.
\end{equation}
A cluster-robust variance estimator is therefore
\begin{equation}
\widehat V_h=\frac{1}{N_C}\sum_{c=1}^{N_C}\hat\IF_{c,h}^2.
\end{equation}
\end{proposition}

\begin{proof}
A first-order mean-value expansion of the moment equations around $\eta_h^0$ gives
\begin{equation}
0=\frac{1}{N_C}\sum_{c=1}^{N_C}\Psi_{c,h}(\eta_h^0)+A_h(\hat\eta_h-\eta_h^0)+o_p(N_C^{-1/2}).
\end{equation}
Rearranging yields
\begin{equation}
\sqrt{N_C}(\hat\eta_h-\eta_h^0)
=
-A_h^{-1}\frac{1}{\sqrt{N_C}}\sum_{c=1}^{N_C}\Psi_{c,h}(\eta_h^0)+o_p(1).
\end{equation}
Applying the delta method to $g(\eta_h)=\mu_{1h}-\mu_{0h}$ gives \eqref{eq:if_linear}.
\end{proof}

Any fixed linear summary of the event-study path inherits an influence function by linearity. For example, if $\HH_+=\{h\in\HH:h\geq 0\}$, then the average post-treatment effect
\begin{equation}
\bar\theta_{post}=\frac{1}{|\HH_+|}\sum_{h\in\HH_+}\theta_h
\end{equation}
has influence function
\begin{equation}
\IF_{c,post}=\frac{1}{|\HH_+|}\sum_{h\in\HH_+}\IF_{c,h}.
\end{equation}
This is the formula used for the pointwise confidence intervals reported in the Monte Carlo table in the main text.

For simultaneous inference on the full event-study vector, the paper uses a multiplier bootstrap based on the stacked estimated influence functions $\hat\Psi_c=(\hat\IF_{c,h})_{h\in\HH}$. Conditional on the sample, draw i.i.d. multipliers $\xi_c$ with mean zero and variance one, and compute
\begin{equation}
\hat\Theta^{\ast}=\hat\Theta+N_C^{-1/2}\sum_{c=1}^{N_C}\xi_c\hat\Psi_c,
\qquad
\hat\Theta=(\hat\theta_h)_{h\in\HH}.
\end{equation}
The simultaneous critical value is the conditional $(1-\alpha)$ quantile of the sup-$t$ statistic
\begin{equation}
T^{\ast}=\max_{h\in\HH}\left|\frac{\hat\theta_h^{\ast}-\hat\theta_h}{\hat\sigma_h}\right|.
\end{equation}
The resulting band is
\begin{equation}
\left[\hat\theta_h-c_{1-\alpha}^{\ast}\hat\sigma_h,\ \hat\theta_h+c_{1-\alpha}^{\ast}\hat\sigma_h\right],
\qquad h\in\HH.
\end{equation}
In practice, the multiplier distribution may be implemented with Rademacher, Mammen, or Webb-type weights. The results in the paper use the multiplier bootstrap for dynamic event-study paths and cluster-robust pointwise analytic inference for the scalar Monte Carlo summaries so as to remain directly comparable with the traditional LP-DiD estimators.

\section{Monte Carlo design}

The Monte Carlo in the main text follows a staggered-adoption panel design with heterogeneous dynamic treatment effects and covariate-driven treatment timing. Its goal is not to compare every possible LP-DiD aggregation rule but to isolate the role of nuisance adjustment in the local ATT problem. That is why the comparison focuses on the unadjusted LP-DiD benchmark, its covariate-adjusted variants, and the proposed DR estimator.

The panel contains $T=17$ periods. Treated units adopt in cohorts $g\in\{9,10,11,12,13,14\}$, and the remaining units are never treated. The main nuisance-misspecification exercise is run at $N\in\{500,250\}$ with 500 replications for each sample size. Let $X=(X_1,X_2,X_3,X_4)'\sim\mathcal N(0,\Sigma)$, with $\Sigma_{kj}=c^{|k-j|}$ and $c=0$ in the baseline design. Define nonlinear transforms $Z=(Z_1,Z_2,Z_3,Z_4)'$ from $X$ by
\begin{align}
\tilde Z_1 &= \exp(0.5X_1), &
\tilde Z_2 &= 10+\frac{X_2}{1+\exp(X_1)}, \\
\tilde Z_3 &= \left(0.6+\frac{X_1X_3}{25}\right)^3, &
\tilde Z_4 &= (20+X_2+X_4)^2,
\end{align}
and standardize componentwise,
\begin{equation}
Z_j=\frac{\tilde Z_j-\E[\tilde Z_j]}{\sqrt{\Var(\tilde Z_j)}},
\qquad j=1,\dots,4.
\end{equation}

Treatment timing is assigned by a multinomial logit over the six treated cohorts plus the never-treated state. For a generic feature vector $W=(W_1,W_2,W_3,W_4)'$, the cohort-$g_j$ logit index is
\begin{equation}
\ell_j(W)=\xi\left(1-\frac{j}{G}\right)\left(-W_1+0.5W_2-0.25W_3-0.2W_4\right),
\qquad j=1,\dots,G,
\end{equation}
with $G=6$ and $\xi=0.9$, while the never-treated state has an index normalized to zero. Drawing from the implied multinomial probabilities determines $G_i$.

For the untreated-outcome process, define the baseline linear index
\begin{equation}
f^{base}_{reg,t}(W)=210+\frac{t}{T}\left(27.4W_1+13.7W_2+13.7W_3+13.7W_4\right).
\end{equation}
In scenarios A and C, the untreated potential outcome uses this baseline index with $W=Z$. In scenarios B and D, the untreated outcome is intentionally made harder through
\begin{align}
\tilde H_1 &= \exp(0.5X_1), &
\tilde H_2 &= X_2^2, \\
\tilde H_3 &= X_2X_3, &
\tilde H_4 &= \sin(X_1+X_4),
\end{align}
followed by componentwise standardization $H_j=(\tilde H_j-\E[\tilde H_j])/\sqrt{\Var(\tilde H_j)}$, and
\begin{equation}
f^{hard}_{reg,t}(X)=210+\frac{t}{T}\left(27.4H_1+13.7H_2+13.7H_3+13.7H_4\right).
\end{equation}
Potential outcomes are then generated as
\begin{align}
Y_{it}(0) &= f_{reg,t} + t + \alpha_i + \delta\,\ind\{G_i<\infty\}\frac{t}{T}(1+0.2X_{i1}) + \varepsilon_{i0t}, \\
Y_{it}(1) &= Y_{it}(0) + \tau_{it},
\end{align}
where $\alpha_i\sim\mathcal N(\mu_i,1)$ with $\mu_i=G_i$ for treated units and $\mu_i=T+1$ for never-treated units, $\varepsilon_{i0t}\sim\mathcal N(0,1)$, and
\begin{equation}
\tau_{it}=\max\{t-G_i+1,0\}(1+0.1X_{i1}).
\end{equation}
Observed outcomes equal $Y_{it}(1)$ once treatment is on and $Y_{it}(0)$ otherwise.

The main nuisance-misspecification family sets $\delta=0$ and differs only in the features used to generate the untreated-outcome and treatment-timing components:
\begin{align}
\text{A: Both nuisances aligned} &\quad (f_{reg,t},W_{ps})=(f^{base}_{reg,t}(Z), Z), \\
\text{B: OR misspecified} &\quad (f_{reg,t},W_{ps})=(f^{hard}_{reg,t}(X), Z), \\
\text{C: PS misspecified} &\quad (f_{reg,t},W_{ps})=(f^{base}_{reg,t}(Z), X), \\
\text{D: Both nuisances misspecified} &\quad (f_{reg,t},W_{ps})=(f^{hard}_{reg,t}(X), X).
\end{align}
The estimating equations always use the observed covariate vector $Z$ in both nuisance steps, so the Monte Carlo isolates misspecification by changing the DGP and not the implementation.

Two implementation details are worth stating explicitly. First, the benchmark LP-DiD coefficient uses the variance-weighted aggregation in Proposition \ref{prop:benchmark_weights}, whereas LPDID-RA, DRLPDID-IPT, and DRLPDID target the treated-entry-weighted object in \eqref{eq:att_target_h}. Second, the pointwise confidence intervals in the main-text table are based on the cluster-robust influence-function standard errors from Proposition \ref{prop:if}. This keeps the simulation directly comparable to the usual LP-DiD reporting convention.

Appendix Table~\ref{tab:sim_main_n250} reports the same Monte Carlo summary for the smaller sample size. The qualitative ranking is similar, although the differences among the adjusted estimators narrow, and finite-sample dispersion increases.

\begin{table}[t!]
\centering
\caption{Monte Carlo: average post-treatment effect ($N=250$)}
\label{tab:sim_main_n250}
\footnotesize
\begin{tabular}{llccc}
\toprule
Scenario & Estimator & Bias & RMSE & Coverage \\
\midrule
\multirow{5}{*}{\shortstack[l]{Both nuisances\\aligned}}
  & LPDID-RW      & -2.352 & 2.817 & 0.702 \\
  & LPDID-RW + X  &  0.067 & 0.130 & 0.914 \\
  & LPDID-RA      &  0.006 & 0.109 & 0.964 \\
  & DRLPDID-IPT   & -0.207 & 0.251 & 0.746 \\
  & DRLPDID       &  0.008 & 0.118 & 0.956 \\
\midrule
\multirow{5}{*}{OR misspecified}
  & LPDID-RW      & -3.632 & 3.950 & 0.384 \\
  & LPDID-RW + X  & -0.016 & 0.842 & 0.922 \\
  & LPDID-RA      & -0.084 & 0.830 & 0.930 \\
  & DRLPDID-IPT   & -0.211 & 0.825 & 0.892 \\
  & DRLPDID       &  0.124 & 0.924 & 0.870 \\
\midrule
\multirow{5}{*}{PS misspecified}
  & LPDID-RW      & -1.701 & 2.307 & 0.830 \\
  & LPDID-RW + X  &  0.066 & 0.130 & 0.908 \\
  & LPDID-RA      & -0.003 & 0.107 & 0.956 \\
  & DRLPDID-IPT   & -0.160 & 0.215 & 0.844 \\
  & DRLPDID       & -0.003 & 0.122 & 0.934 \\
\midrule
\multirow{5}{*}{\shortstack[l]{Both nuisances\\misspecified}}
  & LPDID-RW      & -3.649 & 3.947 & 0.374 \\
  & LPDID-RW + X  & -0.124 & 0.852 & 0.920 \\
  & LPDID-RA      & -0.180 & 0.845 & 0.928 \\
  & DRLPDID-IPT   & -0.324 & 0.840 & 0.880 \\
  & DRLPDID       &  0.017 & 0.889 & 0.894 \\
\bottomrule
\end{tabular}
\vspace{0.25em}

\parbox{0.92\linewidth}{\footnotesize Notes: All entries refer to the average post-treatment effect. Results are based on 500 Monte Carlo replications, with pointwise cluster-robust confidence intervals.}
\end{table}

\subsection{Sensitivity to violations of local conditional parallel trends}

The appendix also reports a supplementary stress test that perturbs the untreated outcome of ever-treated units through the drift term indexed by $\delta$. We consider $\delta\in\{0,0.5,1.0\}$ while keeping the nuisance-aligned specification $(f^{base}_{reg,t}(Z),Z)$. Table~\ref{tab:pt_sensitivity} summarizes the average post-treatment effect for this exercise.

\begin{table}[t!]
\centering
\caption{Sensitivity to violations of local conditional parallel trends}
\label{tab:pt_sensitivity}
\footnotesize
\begin{tabular}{llccc}
\toprule
Scenario & Estimator & Bias & RMSE & Coverage \\
\midrule
\multicolumn{5}{l}{\textit{Panel A.} $N=500$} \\
\midrule
\multirow{5}{*}{$\delta=0$}
  & LPDID-RW      & -2.436 & 2.680 & 0.440 \\
  & LPDID-RW + X  &  0.052 & 0.093 & 0.900 \\
  & LPDID-RA      & -0.008 & 0.077 & 0.964 \\
  & DRLPDID-IPT   & -0.215 & 0.238 & 0.470 \\
  & DRLPDID       & -0.010 & 0.079 & 0.966 \\
\midrule
\multirow{5}{*}{$\delta=0.5$}
  & LPDID-RW      & -2.373 & 2.601 & 0.464 \\
  & LPDID-RW + X  &  0.170 & 0.189 & 0.454 \\
  & LPDID-RA      &  0.107 & 0.134 & 0.736 \\
  & DRLPDID-IPT   & -0.121 & 0.156 & 0.820 \\
  & DRLPDID       &  0.107 & 0.137 & 0.754 \\
\midrule
\multirow{5}{*}{$\delta=1.0$}
  & LPDID-RW      & -2.241 & 2.492 & 0.504 \\
  & LPDID-RW + X  &  0.257 & 0.269 & 0.118 \\
  & LPDID-RA      &  0.198 & 0.213 & 0.332 \\
  & DRLPDID-IPT   & -0.042 & 0.107 & 0.954 \\
  & DRLPDID       &  0.192 & 0.209 & 0.366 \\
\midrule
\multicolumn{5}{l}{\textit{Panel B.} $N=250$} \\
\midrule
\multirow{5}{*}{$\delta=0$}
  & LPDID-RW      & -2.489 & 3.005 & 0.684 \\
  & LPDID-RW + X  &  0.060 & 0.130 & 0.926 \\
  & LPDID-RA      &  0.002 & 0.113 & 0.952 \\
  & DRLPDID-IPT   & -0.218 & 0.269 & 0.728 \\
  & DRLPDID       &  0.002 & 0.118 & 0.948 \\
\midrule
\multirow{5}{*}{$\delta=0.5$}
  & LPDID-RW      & -2.344 & 2.840 & 0.706 \\
  & LPDID-RW + X  &  0.167 & 0.206 & 0.688 \\
  & LPDID-RA      &  0.106 & 0.158 & 0.850 \\
  & DRLPDID-IPT   & -0.126 & 0.201 & 0.890 \\
  & DRLPDID       &  0.097 & 0.161 & 0.856 \\
\midrule
\multirow{5}{*}{$\delta=1.0$}
  & LPDID-RW      & -2.376 & 2.822 & 0.714 \\
  & LPDID-RW + X  &  0.259 & 0.283 & 0.412 \\
  & LPDID-RA      &  0.194 & 0.223 & 0.618 \\
  & DRLPDID-IPT   & -0.054 & 0.157 & 0.950 \\
  & DRLPDID       &  0.189 & 0.224 & 0.634 \\
\bottomrule
\end{tabular}
\vspace{0.25em}

\parbox{0.92\linewidth}{\footnotesize Notes: Supplementary sensitivity exercise based on 500 Monte Carlo replications per design, with pointwise cluster-robust confidence intervals. The DGP keeps the nuisance-aligned specification $(f^{base}_{reg,t}(Z),Z)$ and perturbs the untreated outcome of ever-treated units through the drift parameter $\delta$.}
\end{table}

The sensitivity exercise now confirms, with a full 500-replication design, the expected limit of the method: once local conditional parallel trends is deliberately perturbed, all estimators except the unadjusted RW benchmark move away from the target in the positive direction, and coverage declines sharply as $\delta$ rises. At $\delta=0$, DRLPDID and LPDID-RA are essentially unbiased with near-nominal coverage, whereas DRLPDID-IPT exhibits a negative bias and substantial undercoverage even though the stress parameter is absent. This pattern points to an estimand gap rather than an inference failure in the pure IPT version: supplementary diagnostics show that its cluster-robust standard errors track the empirical dispersion closely, but the pure odds reweighting does not by itself recover the treated-entry-weighted cohort-size target that defines $\theta_h$. As $\delta$ increases, the apparent improvement of DRLPDID-IPT should therefore not be read as robustness to violations of local parallel trends; it reflects offsetting biases of opposite sign. The practical implication is the same as in the main Monte Carlo: use DRLPDID rather than its IPT-only counterpart when covariates matter for local selection and untreated dynamics.

\section{Additional empirical output: richer covariate specification}

The main text reports the parsimonious adjusted application with \texttt{asmrh} only. This appendix reports the same exercise under the richer covariate set \texttt{pcinc}+\texttt{asmrh}+\texttt{cases}, using the canonical no-fault-divorce state-year panel studied by \citet{StevensonWolfers2006} and later reanalyzed by \citet{GoodmanBacon2021}.

The data form an annual panel for the 48 continental U.S. states plus Washington, D.C., from 1964 to 1996. The outcome is \texttt{asmrs}, the female suicide mortality rate. Treatment is \texttt{post}, an indicator equal to one once a state's no-fault-divorce reform is in force and zero beforehand. The richer control set follows the original empirical application: \texttt{pcinc} denotes per-capita income, \texttt{asmrh} is the homicide mortality rate, and \texttt{cases} is the divorce-case filing measure. We treat this richer specification as a supplementary robustness exercise rather than as the preferred adjusted design because some of these variables may be more closely connected to institutional or behavioral margins affected by the reform itself.

\begin{table}[!htbp]
\centering
\caption{Average post-treatment effects in the no-fault-divorce application: richer specification}
\label{tab:app_att_allx_v3}
\small
\begin{tabular}{lrrrr}
\toprule
Estimator & ATT & SE & CI lower & CI upper \\
\midrule
DRLPDID                   & -5.625 & 3.966 & -13.397 &  2.148 \\
LPDID-RW + X              & -8.931 & 3.380 & -15.555 & -2.307 \\
LPDID-RA                  & -6.137 & 3.445 & -12.889 &  0.615 \\
Callaway--Sant'Anna       & -3.416 & 3.240 & -10.239 &  2.744 \\
Borusyak--Jaravel--Spiess & -5.364 & 2.612 & -10.714 & -0.623 \\
Sun--Abraham              & -7.732 & 3.308 & -13.905 & -1.283 \\
Gardner                   & -5.364 & 2.852 & -10.931 &  0.059 \\
\bottomrule
\end{tabular}
\begin{minipage}{0.92\textwidth}
\vspace{0.4em}
\footnotesize
\emph{Note.} Estimates are taken directly from the richer-specification application output. LPDID-RW + X reports the covariate-adjusted local LP-DiD comparator from the Dube-family implementation. LPDID-RA reports the regression-adjusted local estimator from the semiparametric implementation. External staggered-adoption comparators are computed in Python using the \texttt{diff-diff} package \citep{diffdiff2026}.
\end{minipage}
\end{table}

\begin{table}[!htbp]
\centering
\caption{DRLPDID-IPT in the no-fault-divorce application: richer specification}
\label{tab:app_att_allx_ipw}
\small
\begin{tabular}{lrrrr}
\toprule
Estimator & ATT & SE & CI lower & CI upper \\
\midrule
DRLPDID-IPT & -7.262 & 3.050 & -13.241 & -1.283 \\
\bottomrule
\end{tabular}
\begin{minipage}{0.92\textwidth}
\vspace{0.4em}
\footnotesize
\emph{Note.} This table is reported separately only as an additional comparison for the semiparametric family. It is not part of the main cross-estimator application table used in the article.
\end{minipage}
\end{table}

For completeness, the IPT-only version of the semiparametric local estimator is more negative in the richer specification, with an average post-treatment effect of $-7.26$. This places it closer to Sun--Abraham than to DRLPDID, LPDID-RW + X, or LPDID-RA, reinforcing the broader Monte Carlo message that the IPT-only version is the most sensitive member of the semiparametric family.

\begin{figure}[!htbp]
    \centering
    \includegraphics[width=0.95\textwidth]{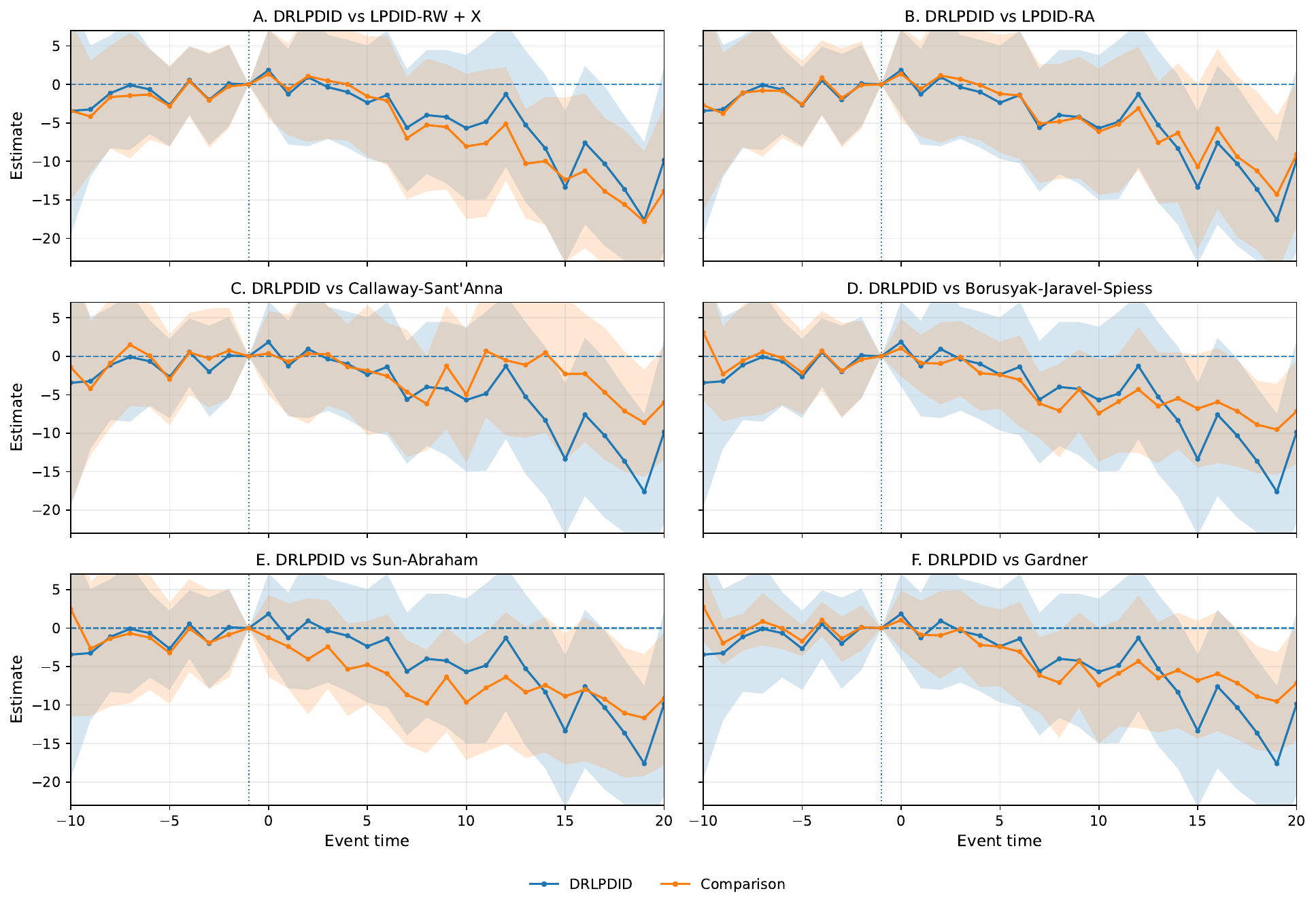}
    \caption{Event-study comparison in the no-fault-divorce application: richer specification}
    \label{fig:app_allx_grid_v3}
\end{figure}

Table~\ref{tab:app_att_allx_v3} and Figure~\ref{fig:app_allx_grid_v3} show that the richer specification does not overturn the main empirical ranking. DRLPDID yields $-5.62$, again close to Borusyak--Jaravel--Spiess and Gardner (both $-5.36$). LPDID-RA is somewhat more negative at $-6.14$, whereas LPDID-RW + X remains substantially more negative at $-8.93$. Callaway--Sant'Anna becomes noticeably less negative at $-3.42$, while Sun--Abraham remains more negative at $-7.73$. The richer specification, therefore, serves as a robustness diagnostic rather than as the headline empirical result.